\documentclass[11pt,aps,
]{revtex4}
\usepackage{amsmath,amssymb,amsthm,amscd,mathrsfs}
\usepackage{makeidx}
\usepackage{subfigure}
\usepackage{graphicx}
\usepackage{natbib}

\usepackage{xcolor}

\DeclareMathOperator{\sech}{sech}
\DeclareMathOperator{\csch}{csch}

\newcommand{\tcm}{\textcolor{magenta}}

\begin{document}
\title{Quasi-exactly solvable hyperbolic potential and its anti-isospectral counterpart}
\author{E. Condori-Pozo\footnote{E-mail: econdoripozo@fisica.ugto.mx}}
\author{M. A. Reyes\footnote{E-mail: marco@fisica.ugto.mx}}
\affiliation{Departamento de F\'{\i}sica, Universidad de Guanajuato, Apdo. Postal E143 Le\'on, Gto., Mexico}
\author{H.C. Rosu\footnote{E-mail: hcr@ipicyt.edu.mx}}
\affiliation{Instituto Potosino de Investigaci\'on Cient\'{\i}fica y Tecnol\'ogica,\\
Camino a la presa San Jos\'e 2055, Col. Lomas 4a Secci\'on, 78216 San Luis Potos\'{\i}, S.L.P., Mexico}

\begin{abstract}
We solve the eigenvalue spectra for two quasi exactly solvable (QES) Schr\"odinger problems defined by the potentials $V(x;\gamma,\eta) = 4\gamma^{2}\cosh^{4}(x) + V_{1}(\gamma,\eta) \cosh^{2}(x) + \eta \left( \eta-1 \right)\tanh^{2}(x)$
and
$ U(x;\gamma,\eta) = -4\gamma^{2}\cos^{4}(x) - V_{1}(\gamma,\eta)\cos^{2}(x) + \eta \left( \eta-1 \right)\tan^{2}(x)$,
found by the anti-isospectral transformation of the former. We use three methods: a direct polynomial expansion, which shows the relation between the expansion order and the shape of the potential function; direct comparison to the confluent Heun equation (CHE), which has been shown to provide only part of the spectrum in different quantum mechanics problems, and the use of Lie algebras, which has been proven to reveal hidden algebraic structures of this kind of spectral problems. 
\end{abstract}

\maketitle

\section{Introduction} \label{sec:intro}

In quantum mechanics, a QES spectral problem is one for which it is not possible to obtain the complete energy spectrum analytically \citep[]{turbiner:1988, ushveridze:1994}. This class of Schr\"odinger problems is not limited to those of single particles, as there are also examples of QES multiple-body problems in one, two, and higher dimensions \citep[]{khare:1998}. In the literature, there exist three direct methods to find the solutions of QES quantum-mechanical problems: ($i$) use of a simple polynomial expansion, ($ii$) transforming the Schr\"odinger equation into a known equation, like the CHE, and ($iii$) using Lie algebras. The first method is used to show, in a very simple way, that for some QES problems, the wave solutions and the potentials share an intimate relation that preclude to fix the latter and allow for a recursive method to generate the complete spectrum. Secondly, problems where the Heun equation is used to find the spectra have been shown to be in the QES class, and in our case we find that the solutions found in the first method turn out to be the same.  On the other hand, the use of Lie algebraic methods have been used to reveal the existence of hidden algebraic structures in problems which do not show any hidden symmetry properties, a feature which according to Turbiner \citep[]{turbiner:2016} was first noticed by Zaslavskii and Ulyanov \citep[]{zaslavskii:2016}. In the latter framework, the comprehensive review by Turbiner \citep[]{turbiner:2016} emphasises the hidden symmetries involved and its application to finite difference equations. He  was the first to give a detailed list of potentials for QES problems, while Gonz\'alez-L\'opez and collaborators further developed these techniques to a larger list of potentials \citep[]{gonzalez-lopez:1993,artemio:1994,finkel:1996}. However, being simpler, the polynomial expansion based on the Bethe Ansatz method \citep[]{zhang:2012} and the CHE transformation that we also use here have been the common tools in many more recent studies on QES spectral problems \citep[]{finkel:1999,xie:2012,downing:2013,becker:2017,reyes:2018,dong:2019jan,dong:2019feb}.

On the other hand, hyperbolic and trigonometric type potentials are used in molecular physics and quantum
chemistry, modeling inter-atomic and inter-molecular forces, ranging from Razavy \citep[]{razavy:1980}, P\"oschl-Teller \citep[]{teller:1933}, Rosen-Morse \citep[]{morse:1929,morse:1932}, and Scarf \citep[]{scarf:1958} potentials to their modified counterparts \citep[]{suparmi:2019,wu:1990,ikot:2013,gu:2009,zhang:2012,prastyaningrum:2016,schulze:2017}. In quantum chemistry, the area of IR-spectroscopy is particularly interesting, since  double-well potentials (DWP) could describe the ammonia molecule (NH$_3$) \citep[]{sitnitsky:2017}, chromous acid (CrOOH) \citep[]{sitnitsky:2019} and potassium dihydrogen phosphate (KH2PO4) experimental data \citep[]{sitnitsky:2019,lawrence:1980,lawrence:1981}.

For the 1D time independent Schr\"odinger equation
\begin{equation*}
-\frac{d^2\Psi(x)}{dx^2} + V(x)\Psi(x) = E\Psi(x) \ ,
\end{equation*}
with $\frac{\hbar^2}{2m}=1$, we study the spectral problem for the hyperbolic potential
\begin{equation} \label{eq:1}
 V(x;\gamma,\eta) = 4\gamma^{2}\cosh^{4}(x) + V_{1}(\gamma,\eta) \cosh^{2}(x) + \eta \left( \eta-1 \right)\tanh^{2}(x)~,
\end{equation}
and its trigonometric counterpart
\begin{equation} \label{eq:2}
 U(x;\gamma,\eta) = -4\gamma^{2}\cos^{4}(x) - V_{1}(\gamma,\eta)\cos^{2}(x) + \eta \left( \eta-1 \right)\tan^{2}(x)~,
\end{equation}
which resembles the symmetric potential used by Sitnitsky to study the inversion vibrational mode for the ammonia molecule \citep[]{sitnitsky:2017}. In a recent article, Dong {\em et al} \citep[]{dong:2019jan}, using the CHE approach, find analytical solutions for the potential $V(x;c,k) = c^{2} \sinh^{4}(x) - k \tanh^{2}(x)$, which is a particular case of the hyperbolic potential (\ref{eq:1}), when $c$ is written in terms of $\gamma$, $\eta$, and the polynomial order $N$, and $k = - \eta \left( \eta - 1 \right)$. In their work, they only develop the case $k>0$ which corresponds to values $0 < \eta < 1$, but in our case both cases, $k<0$ and $k>0$, are included, making ours a more general treatment.

Furthermore, Sitnitsky \citep[]{sitnitsky:2019} has shown applications to IR spectroscopy for a particle in a DWP
potential to describe the proton energy states in hydrogen bonds. The model potential is a DW trigonometric
potential, with an asymmetric term whose zero limit is used to describe thermodynamic features of the
experimental data.

\medskip

This article is organized as follows.
In Section~\ref{poli}, the direct polynomial expansion solutions are worked out for the quantum spectral problems corresponding to the potentials (\ref{eq:1}) and (\ref{eq:2}).  In Section~\ref{eheun}, we show that the same solutions can be found by a straight comparison to the CHEs, and in Section~\ref{liealg}, we find the explicit Lie algebraic solutions of these problems.  In all cases, we refer to two main trial eigenfunctions that we introduce in Section \ref{poli}, where two additional trial eigenfunctions are worked out separately, due to their non-realization with respect to the anti-isospectral transformation,
$U(x;\gamma,\eta) = - V(ix;\gamma,\eta)$, which relates that part of the spectrum found for the hyperbolic potential with the opposite sign part of the spectrum of the trigonometric potential. The selection of these four trial function types becomes evident in Section IV, where they are related to the four parameter sets of solutions used in previous work by Gonz\'alez-L\'opez {\em et al} \citep[]{finkel:1999}.

\section{Polynomial expansions}\label{poli}

Let us consider the spectral problem for the one-dimensional Schr\"odinger equation
with potentials as given in equations (\ref{eq:1}) and (\ref{eq:2}). We first find the solutions provided by the polynomial expansion of the eigenfunctions. These problems belong to the QES class, and the solvable part of the energy spectrum is found to depend on the order $N$ of the polynomial.

\subsection{The hyperbolic case}\label{ssec21}

In order to use the Bethe ansatz, we work out two classes of even and odd solutions, called here trial functions, {\em TF}.

\paragraph{TF1.}  

We first look for even solutions in the case of the hyperbolic potential (\ref{eq:1}), with eigenfunctions
\begin{equation} \label{eq:3}
    \Psi_{1}(x) = e^{-\gamma \cosh^{2}(x)} \cosh^{\eta}(x) f(x).
\end{equation}
By using the change of variable $z = \cosh^{2}(x)$, we can immediately find the CHE equation for the function $f(z)$,

\begin{align} \label{eq:4}
    \frac{d^{2}f}{dz} + \left[ - 2\gamma + \frac{\eta+\frac{1}{2}}{z} + \frac{\frac{1}{2}}{z-1} \right] \frac{df}{dz} + \left[ \frac{-\frac{1}{4} 
    \left( E + \eta + 2 \gamma \left( 2\eta + 1 \right) \right)}{z} + \frac{\frac{1}{4} 
    \left( E + \eta - V_{1} - 2\gamma \left( 2\gamma + 1 \right) \right)}{z-1} \right]f = 0
\end{align}
and look for the polynomial expansion solution
\begin{equation} \label{eq:5}
    f(z) = f_{0}\prod_{i=1}^{N} \left( z-z_{N,i} \right)~.
\end{equation}

The solutions found in this way depend on the order $N$ of the polynomial: For $N = 0$, we can only find one energy eigenvalue, $E_0 = - \eta-2 \gamma \left(2 \eta + 1\right)$;
for $N = 1$, we find two eigenvalues, $E_{\pm} = -[3\eta + 6 \gamma + 2 + 4\gamma\eta] \pm 2\big[\left( \eta + 1 \right)^{2} + 4\gamma \left( \gamma - \eta \right)\big]^{1/2}$, and the polynomial roots $z_{1,1} = - 2\left( 2\eta + 1 \right)/[E_{\pm}+\eta+2\gamma \left( 2\eta + 1 \right)]$,
while for $N = 2$ we obtain a third order equation for the energy eigenvalues, which we only solve numerically.  With the use of Wolfram's Mathematica, in Table \ref{table:1} we give a summary of the eigenvalues found analytically and numerically, when the parameters are $\gamma=\eta= 2$. In this case,
the polynomial roots are, for $N$=1, $z_{1,1}=0.5,\,1.25$, and for $N$=2, the three pairs of roots are
$(z_{2,1},z_{2,2})=(0.294,\,0.823)$, (0.388,\,1.612), and (1.124,\,2.008).

\begin{figure}[h!]
    \centering
    \subfigure[$\,N$ = 0]{\includegraphics[width=0.32\textwidth]{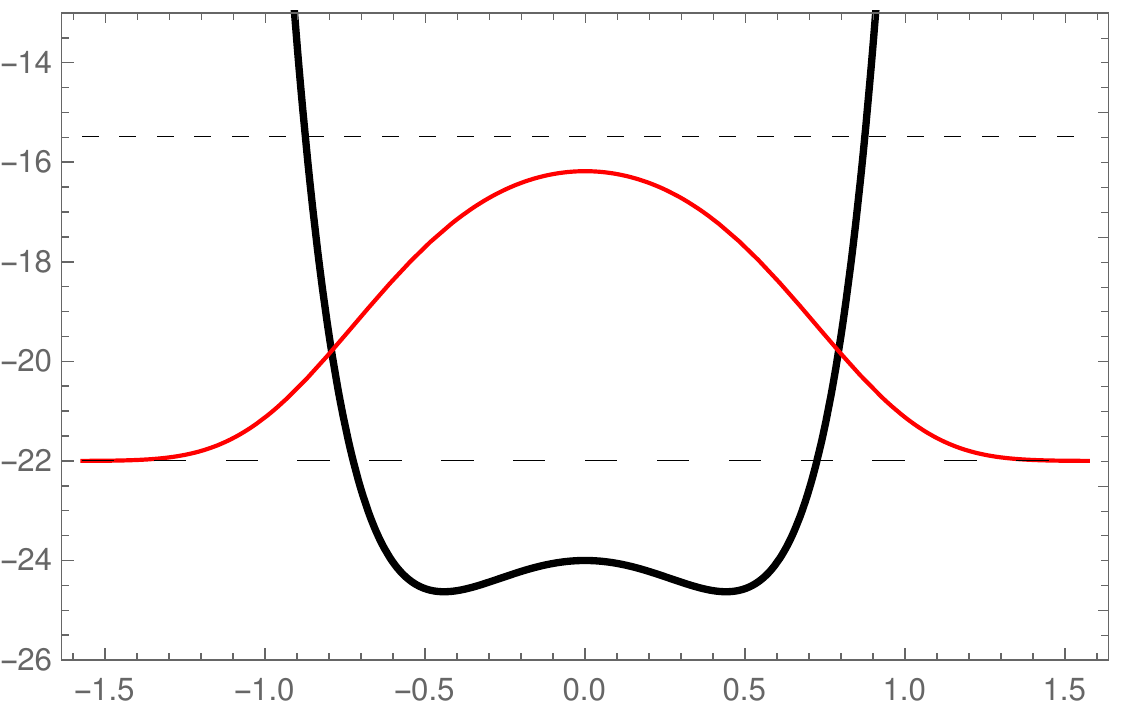}}
    \subfigure[$\,N$ = 1]{\includegraphics[width=0.32\textwidth]{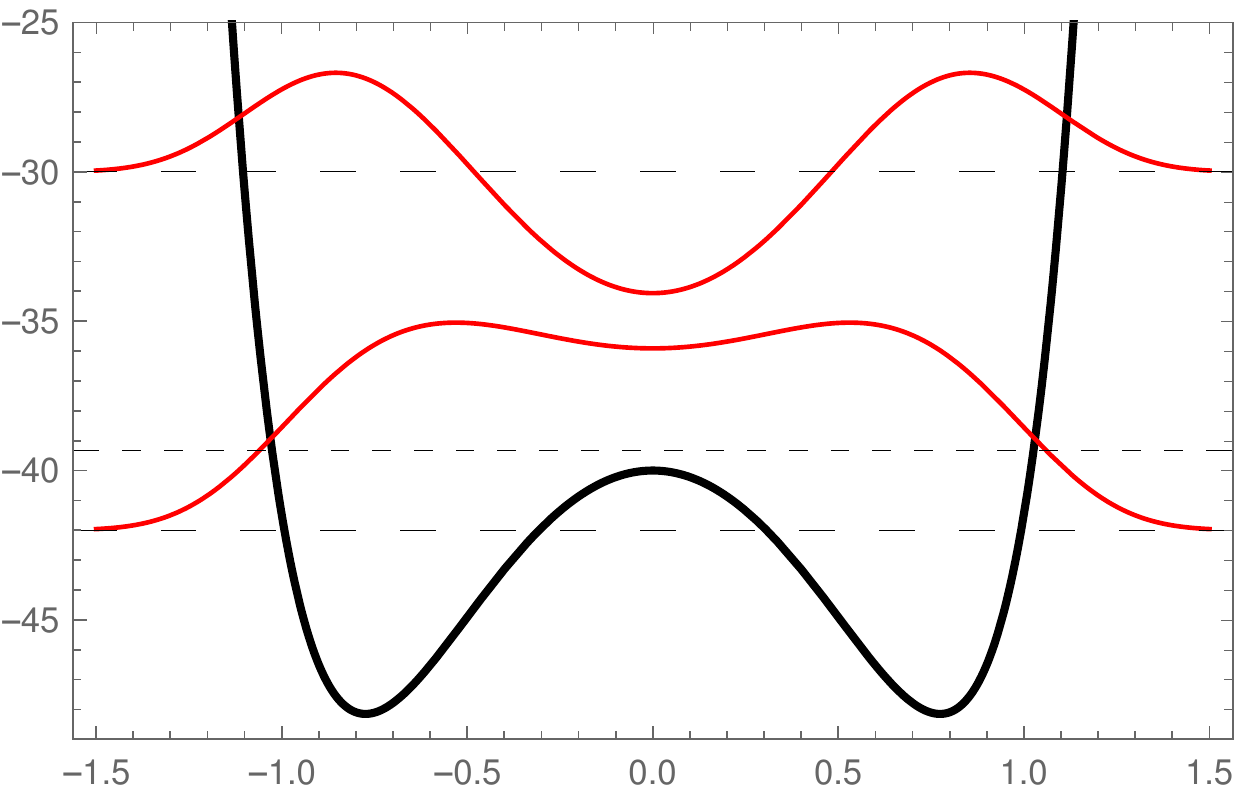}}
    \subfigure[$\,N$ = 2]{\includegraphics[width=0.32\textwidth]{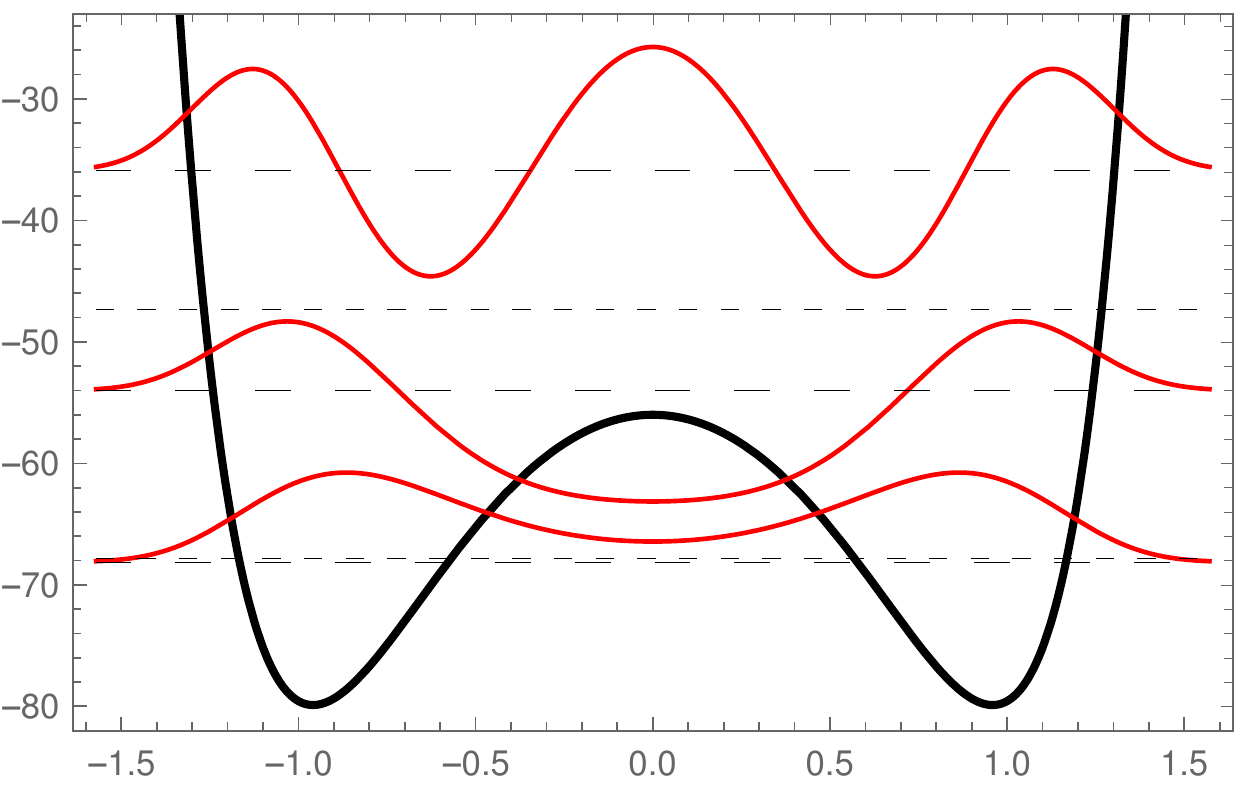}}
\caption{
Hyperbolic potentials (\ref{eq:1}), with $\gamma=2$ and $\eta=2$, and corresponding {\em TF1} eigenfunctions.
}
\label{fig:1}
\end{figure}

In all of these cases, the coefficient $V_1$ is found to be $V_1=-8\gamma[N+1+(\gamma+\eta-1)/2]$, whose dependence on $N$ forbids to scale the solutions of different polynomial expansion orders. Besides, the polynomial order gives the number of roots in each case, and that fixes to $N+1$ the number of eigenfunctions. Three hyperbolic potentials with their analytic wave functions corresponding to this case are plotted in Fig.~\ref{fig:1}, and several
analytical and numerical eigenvalues are provided in Table~\ref{table:1}.

\begin{table}[htp!]
\begin{center}
\begin{tabular}{|c|c|c|c|c|c|c|}
\hline
& \multicolumn{3}{c|}{{\em TF1}} & \multicolumn{3}{c|}{{\em TF2}} \\
\hline
  & $N$ = 0 & $N$ = 1 & $N$ = 2 & $N$ = 0 & $N$ = 1 & $N$ = 2 \\
\hline
$E_{0}$ & $\boldsymbol{-22.000}$ & $\boldsymbol{-42.000}$ & $\boldsymbol{-68.124}$ & $-31.606$ & $-53.922$ & $-84.704$ \\
$E_{1}$ & $-15.489$ & $-39.323$ & $-67.801$ & $\boldsymbol{-27.000}$ & $\boldsymbol{-52.798}$ & $\boldsymbol{-84.635}$ \\
$E_{2}$ & $-5.186$ & $\boldsymbol{-30.000}$ & $\boldsymbol{-54.000}$ &  $-17.502$ & $-42.265$ & $-65.806$ \\
$E_{3}$ &  $7.489$ & $-19.350$ & $-47.331$ &  $-5.773$ & $\boldsymbol{-33.202}$ & $\boldsymbol{-61.915}$ \\
$E_{4}$ & $22.215$ & $-6.315$ & $\boldsymbol{-35.875}$ &  $8.108$ & $-21.011$ & $-50.642$ \\
$E_{5}$ & $38.772$ & $8.674$ & $-22.557$ & $23.880$ & $-6.822$ & $\boldsymbol{-38.449}$ \\
$E_{6}$ & $57.008$ & $25.435$ & $-7.300$ & $41.377$ & $9.198$ & $-24.001$ \\
$E_{7}$ & $76.809$ & $43.837$ & $9.690$ & $60.477$ & $26.900$ & $-7.753$ \\
$\vdots$ & $\vdots$ & $\vdots$ & $\vdots$
 & $\vdots$ & $\vdots$ & $\vdots$ \\
\hline
\end{tabular}
\caption{Exact (bold type) and numerical eigenvalues for the hyperbolic potential (\ref{eq:1}), for even ({\em TF1}) and odd ({\em TF2})
eigenfunctions, with $N$ = $\{0,1,2\}$,
and for $\gamma=\eta=2$; these numerical values are also used in subsequent tables II and III and the figures of the paper.}
\label{table:1}
\end{center}
\end{table}

\paragraph{TF2.} 

\begin{figure}[h!]
    \centering
    \subfigure[ $\,N$ = 0]{\includegraphics[width=0.32\textwidth]{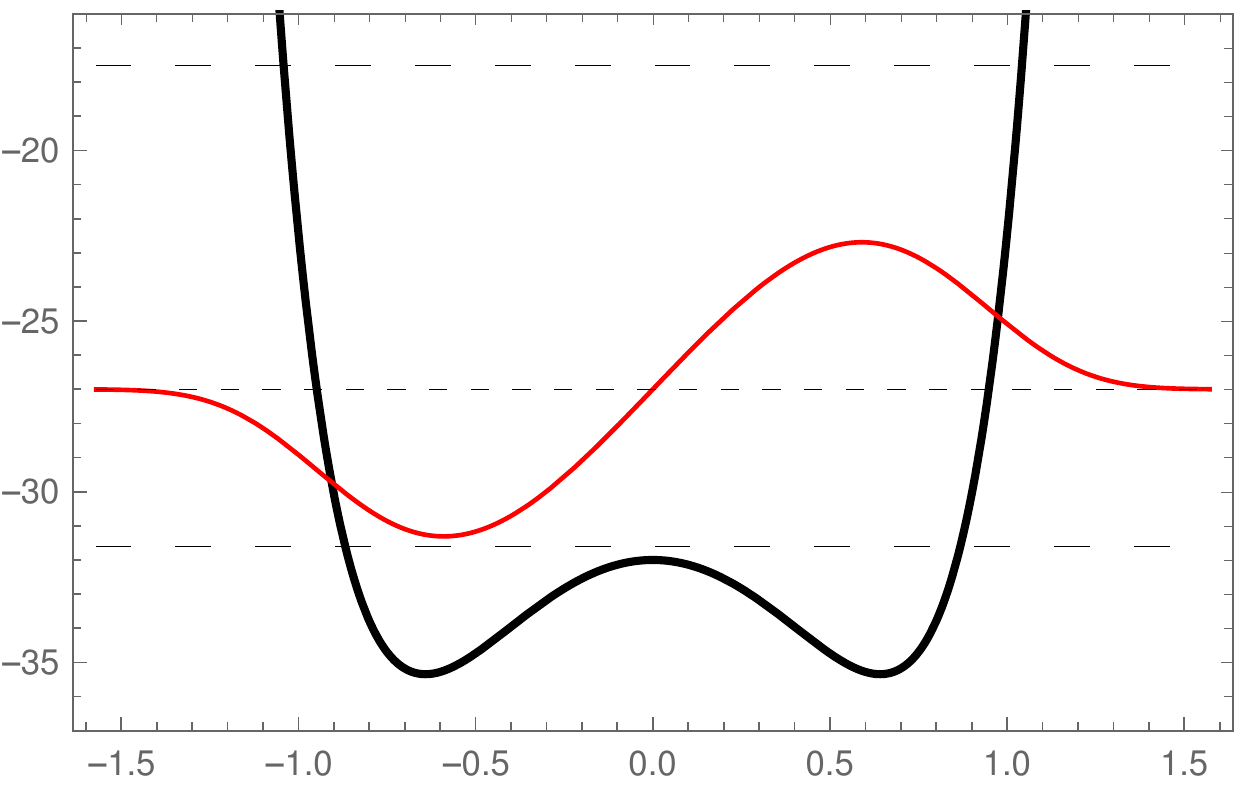}}
    \subfigure[ $\,N$ = 1]{\includegraphics[width=0.32\textwidth]{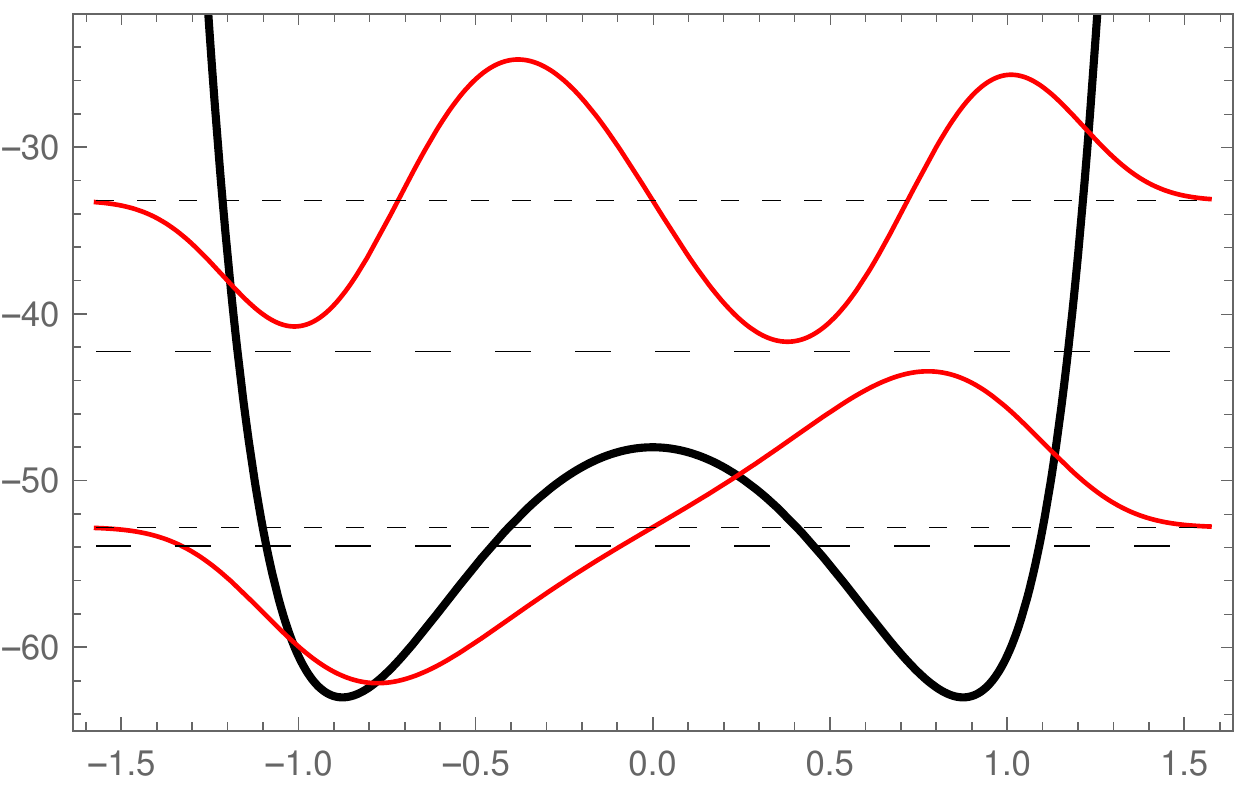}}
    \subfigure[ $\,N$ = 2]{\includegraphics[width=0.32\textwidth]{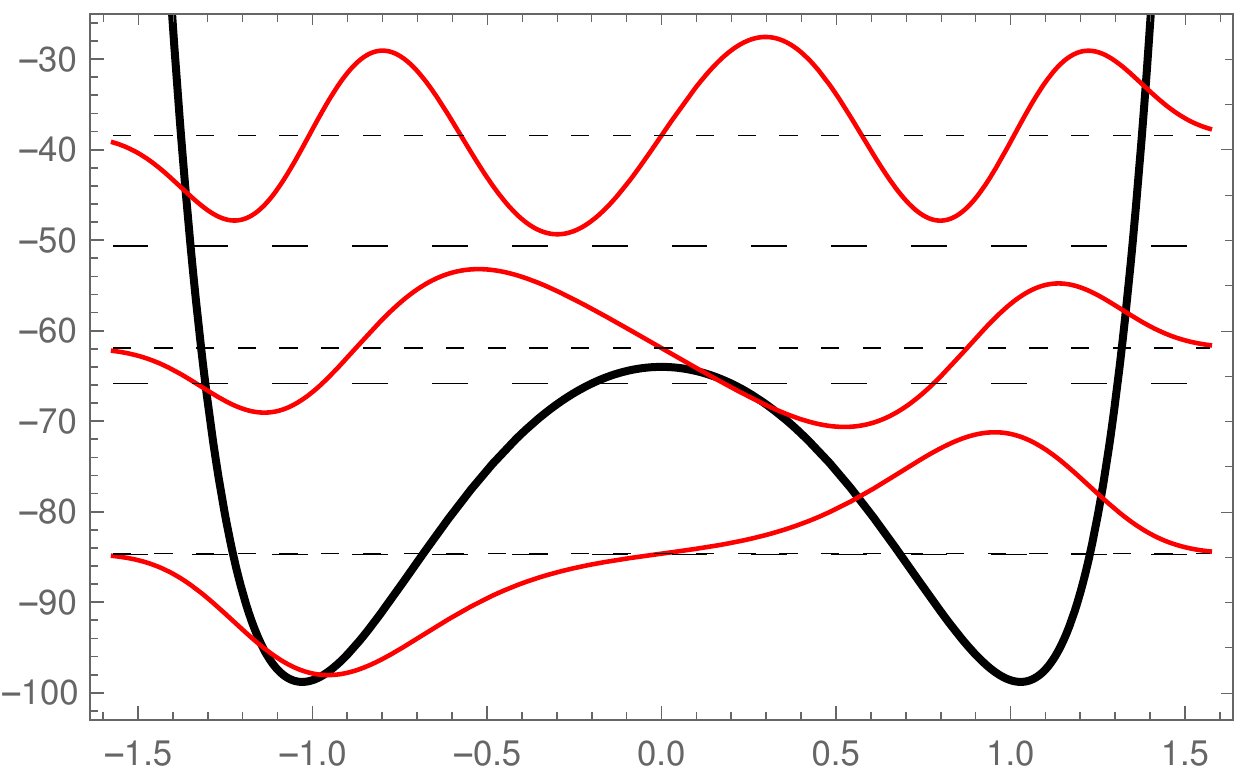}}
\caption{The same as in the previous figure, but 
for the {\em TF2} case.}
\label{fig:2}
\end{figure}

We now look for odd solutions of the type

\begin{equation} \label{eq:6}
    \Psi_{2}(x) = e^{-\gamma \cosh^{2}(x)} \cosh^{\eta}(x) \sinh(x) f(x).
\end{equation}
The CHE for the function $f$ of variable $z = \cosh^{2}(x)$ is given by
{\small
\begin{equation} \label{eq:7}
    \frac{d^{2}f}{dz^{2}} + \left[ -2\gamma + \frac{\eta + \frac{1}{2}}{z} + \frac{\frac{3}{2}}{z-1} \right] \frac{df}{dz} + \left[ \frac{ -
     \frac{1}{4}\left( E + 3\eta +2\gamma \left( 2\eta + 1 \right)+1 
     \right)}{z} + \frac{%
     \frac{1}{4}\left( E+ 3\eta - V_{1} - 2\gamma \left( 2\gamma + 3 \right) + 1 \right)}{z-1} \right] f = 0~.
\end{equation}
}
Again, using the polynomial expansion (\ref{eq:5}), we find that $V_1=-8\gamma[N+1+(\gamma+\eta)/2]$.
Regarding the eigenvalues, for $N = 0$, we get $E_0 = -\eta - (2\gamma +1)(2\eta + 1)$.
For $N = 1$, we find the eigenvalues $E_{\pm} = -[6\gamma + 5\eta + 4\gamma\eta + 5] \pm 2\big[ \left( \eta + 2 \right)^{2} + 4\gamma \left( \gamma - \eta + 1 \right)\big]^{1/2}$, and the roots $z_{1,1} = -2 \left( 2\eta + 1 \right)/\left(E_{\pm} + 2\gamma \left( 2\eta + 1 \right) + 3\eta + 1\right)$, which for $\gamma=2$ and $\eta=2$ are $z_{1,1}=0.388$ and $z_{1,2}=1.612$.
Finally, for the case $N = 2$, with $\gamma = 2$ and $\eta = 2$, the three eigenvalues are
$E_{1} = -84.635$, $E_{3} = -61.915$ and $E_{5} = -38.449$, and the pairs of roots are
$(z_{2,1},z_{2,2})$=(0.235,0.335), (2.417,0.663), and (1.983,1.368).
We plot the potentials and wave functions in Fig.~\ref{fig:2}, and give the analytical and numerical eigenvalues in Table \ref{table:1}.

\paragraph{TF3.} 

For the first two sets of trial functions, the cases with $\eta<0$ are not forbidden, however, we consider them separately due to the restriction that will appear when the anti-isospectral condition is applied, when we turn to the trigonometric case.  So, let us consider here the case of
\begin{equation} \label{eq:8}
    \Psi_{3}(x) = e^{-\gamma \cosh^{2}(x)} \sech^{\eta-1}(x)f(x).
\end{equation}
The exponent of sech($x$) is set for convenience, as will be used in Section \ref{ssec41}.

Using $z = \cosh^{2}(x)$, we arrive at the CHE equation
{\small
\begin{equation} \label{eq:9}
    \frac{d^{2}f}{dz^{2}} + \left[ -2\gamma + \frac{-\eta + \frac{3}{2}}{z} + \frac{\frac{1}{2}}{z-1} \right] \frac{df}{dz} + \left[ \frac{ -\frac{1}{4}
    \left( E- \eta + 2\gamma \left( 3 - 2\eta \right) + 1 \right)}{z} + \frac{\frac{1}{4} \left( E - \eta - V_{1} - 2\gamma \left( 2\gamma + 1 \right) + 1 \right)}{z-1} \right] f = 0~.
\end{equation}
}
The polynomial solutions of order $N$ render the coefficient $V_1=-8\gamma[N+1+(\gamma-\eta)/2]$.
The analytical and numerically found eigenvalues are given in Table \ref{table:2}, for $N$ = $0$, $1$, and $2$.

\begin{table}[htp!]
\begin{center}
\begin{tabular}{|c|c|c|c|c|c|c|}
\hline
& \multicolumn{3}{c|}{{\em TF3}} & \multicolumn{3}{c|}{{\em TF4}} \\
\hline
  & $N$ = 0 & $N$ = 1 & $N$ = 2 & $N$ = 0 & $N$ = 1 & $N$ = 2 \\
\hline
$E_{0}$ & $\boldsymbol{5.000}$ & $\boldsymbol{-12.798}$ & $\boldsymbol{-31.606}$ & $-3.826$ & $-22.000$ & $-42.000$ \\
$E_{1}$ & $16.250$ & $-4.544$ & $-27.000$ & $\boldsymbol{6.000}$ & $\boldsymbol{-15.489}$ & $\boldsymbol{-39.323}$ \\
$E_{2}$ & $29.800$ & $\boldsymbol{6.798}$ & $\boldsymbol{-17.502}$ &  $18.447$ & $-5.186$ & $-30.000$ \\
$E_{3}$ &  $45.329$ & $20.417$ & $-5.773$ &  $33.021$ & $\boldsymbol{7.489}$ & $\boldsymbol{-19.350}$ \\
$E_{4}$ & $62.635$ & $35.998$ & $\boldsymbol{8.108}$ &  $49.464$ & $22.215$ & $-6.315$ \\
$E_{5}$ & $81.579$ & $53.346$ & $23.880$ & $67.610$ & $38.772$ & $\boldsymbol{8.674}$ \\
$E_{6}$ & $102.057$ & $72.325$ & $41.377$ & $87.337$ & $57.008$ & $25.435$ \\
$E_{7}$ & $123.986$ & $92.833$ & $60.477$ & $108.555$ & $76.809$ & $43.837$ \\
$\vdots$ & $\vdots$ & $\vdots$ & $\vdots$
 & $\vdots$ & $\vdots$ & $\vdots$ \\
\hline
\end{tabular}
\caption{Exact (bold type) and numerical eigenvalues for the hyperbolic potential (\ref{eq:1}), for even {\em TF3} and odd {\em TF4} eigenfunctions.}
\label{table:2}
\end{center}
\end{table}

\paragraph{TF4.} 

As said above, we may consider a fourth case, of odd eigenfunctions,

\begin{equation} \label{eq:10}
    \Psi_{4}(x) = e^{-\gamma \cosh^{2}(x)} \sech^{\eta-1}(x)\sinh(x)f(x)~,
\end{equation}
which leads to the CHE
{\small
\begin{equation} \label{eq:11}
    \frac{d^{2}f}{dz^{2}} + \left[ -2\gamma + \frac{-\eta + \frac{3}{2}}{z} + \frac{\frac{3}{2}}{z-1} \right] \frac{df}{dz} + \left[ \frac{ -\frac{1}{4} \left( E - 3\eta + 2\gamma \left( 3 - 2\eta \right) + 4 \right)}{z} + \frac{\frac{1}{4} \left( E- 3\eta - V_{1} - 2\gamma \left( 2\gamma + 3 \right) + 4 \right)}{z-1} \right] f = 0~.
\end{equation}
}
For this case, $V_{1} = -8 \gamma [N+1+(\gamma-\eta+1)/2]$ and the analytical and numerical eigenvalues are given in
Table \ref{table:2}, for the same three values of $N$.

\subsection{The trigonometric case}\label{ssec22}

The trigonometric potential (\ref{eq:2}), can be obtained from the hyperbolic one, eq.~(\ref{eq:1}), via
the anti-isospectral transformation $x \to ix$.  This in turn implies that the trigonometric eigenvalues should be of opposite sign to those of the hyperbolic cases.

\begin{figure}[h!]
    \centering
    \subfigure[ $\,N$ = 0]{\includegraphics[width=0.32\textwidth]{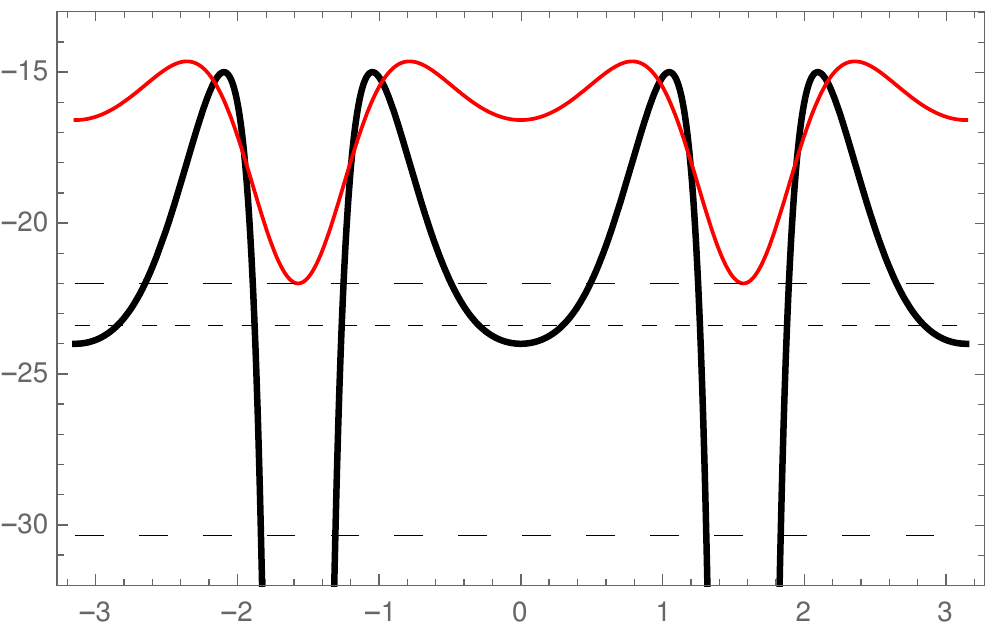}}
    \subfigure[ $\,N$ = 1]{\includegraphics[width=0.32\textwidth]{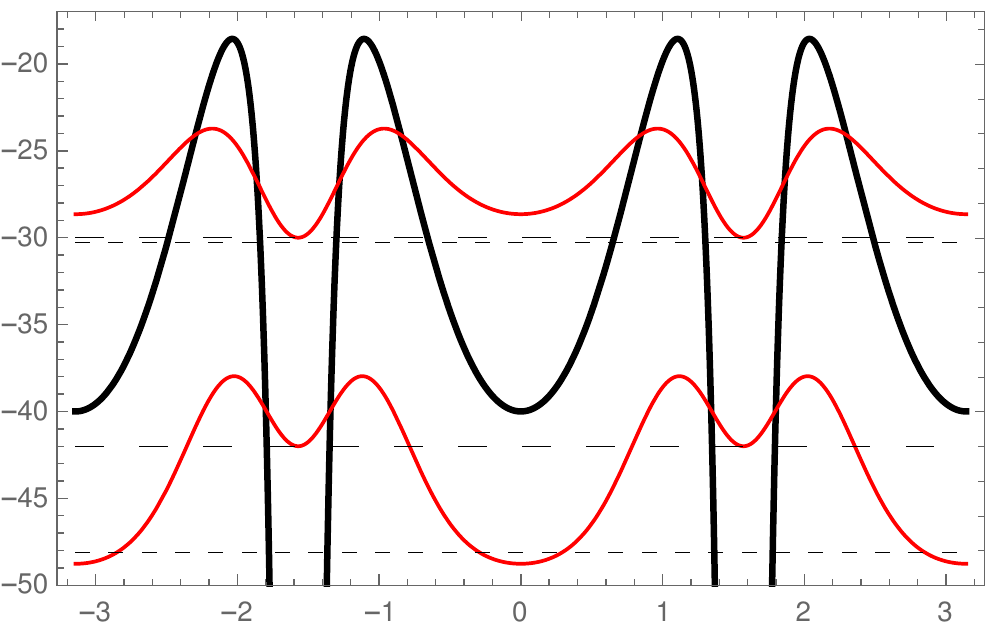}}
    \subfigure[ $\,N$ = 2]{\includegraphics[width=0.32\textwidth]{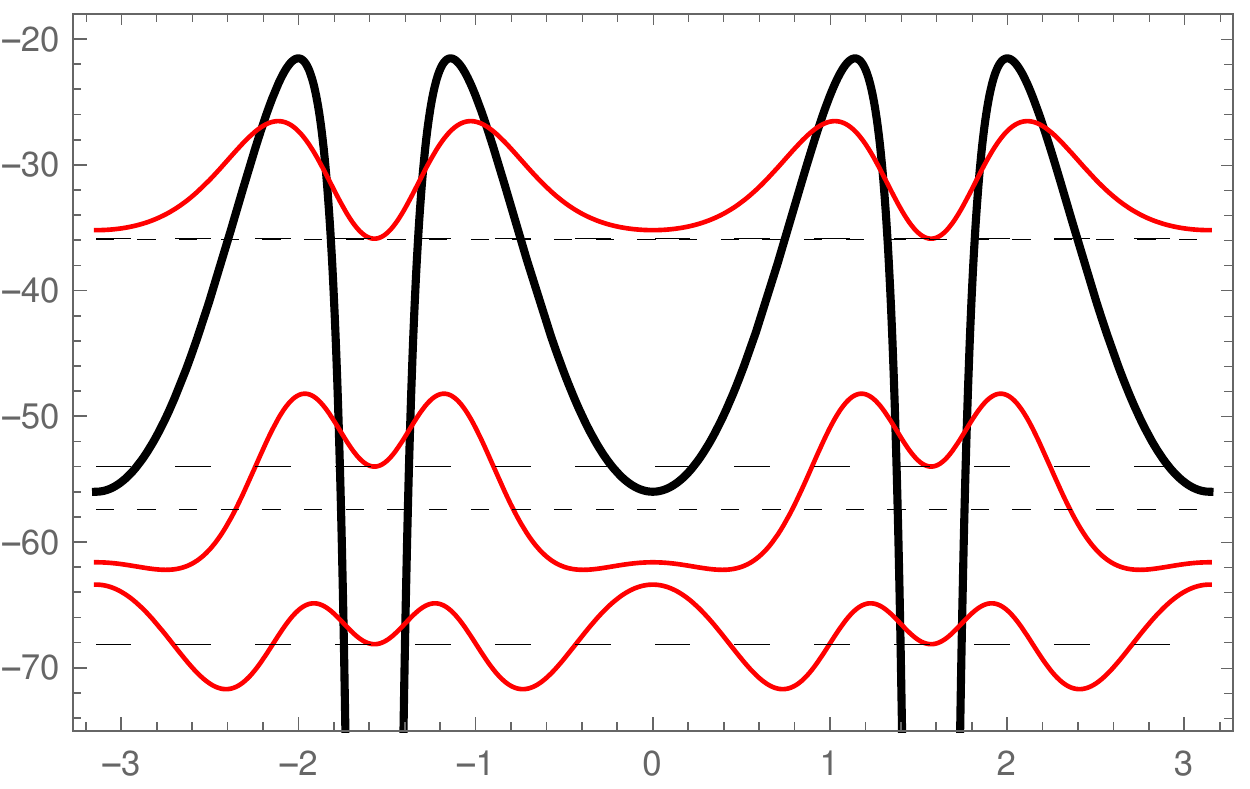}}
\caption{
Negative trigonometric potentials, $-U(x;\gamma = 2,\eta = 2)$, and corresponding {\em TF1} wave functions in one period.}
\label{fig:3}
\end{figure}

\begin{figure}[h!]
    \centering
    \subfigure[ $\,N$ = 0]{\includegraphics[width=0.32\textwidth]{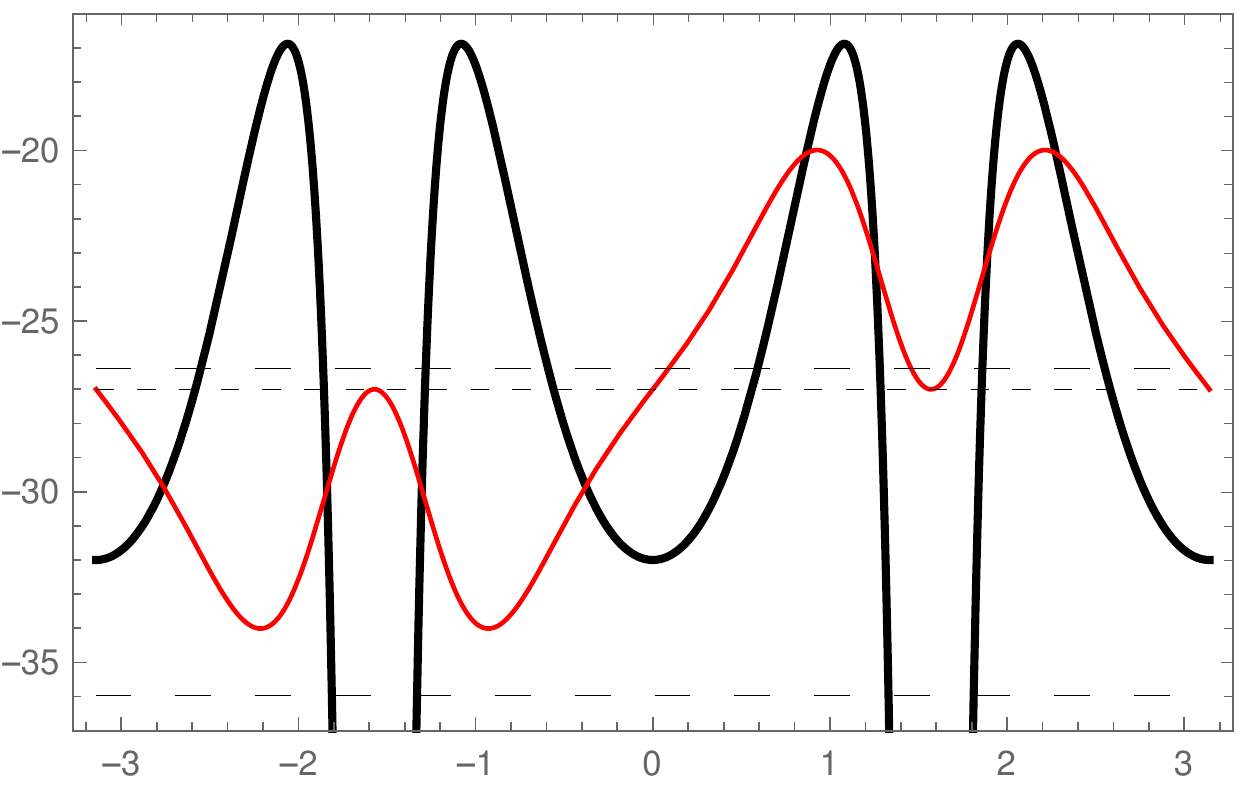}}
    \subfigure[ $\,N$ = 1]{\includegraphics[width=0.32\textwidth]{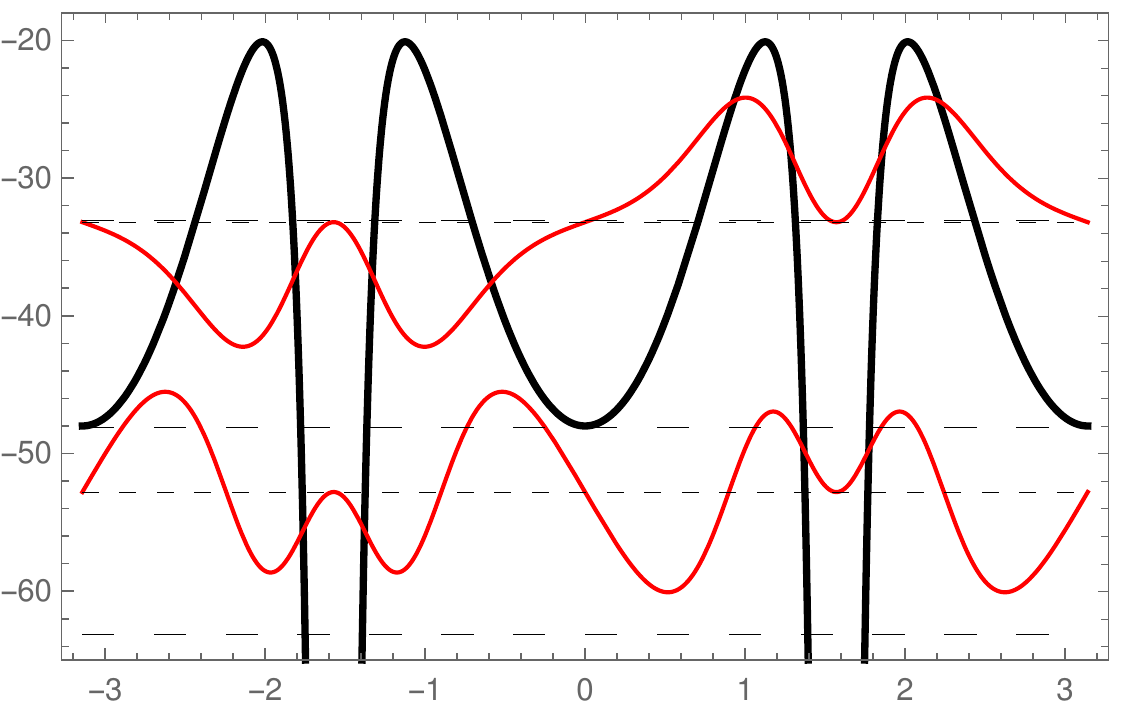}}
    \subfigure[ $\,N$ = 2]{\includegraphics[width=0.32\textwidth]{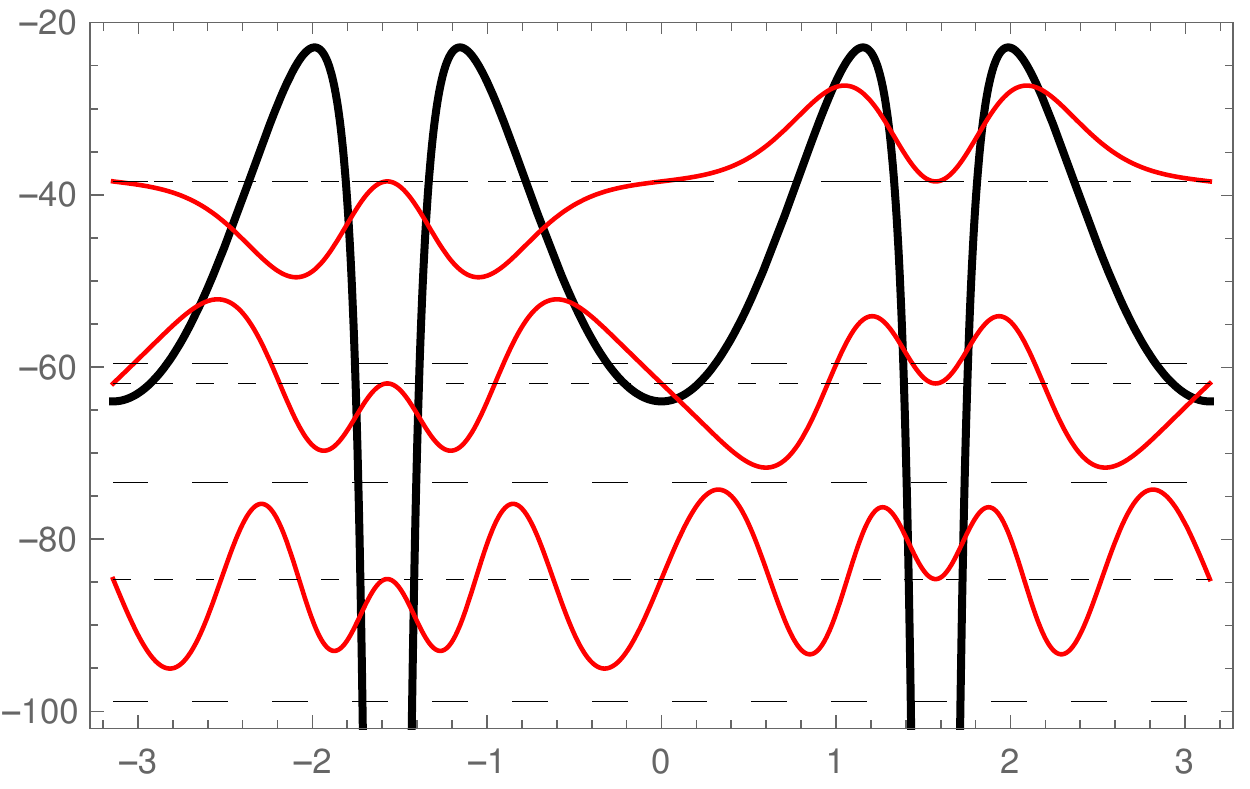}}
\caption{
The same as in the previous figure, but for the {\em TF2} case.} 
\label{fig:4}
\end{figure}

Now we can see that the trial functions {\em TF1} and {\em TF2} can be transformed into regular solutions
of the potential (\ref{eq:2}), while the trial functions {\em TF3} and {\em TF4} would not render square
integrable eigenfunctions due to the sec($x$) term occurring there.  Therefore, the trial functions {\em TF1} and {\em TF2}, for
$\eta>0$, are the only two which can be used in the trigonometric case.  Here we shall only summarize the results for these functions and present
the corresponding plots in Figs.~\ref{fig:3} and \ref{fig:4}.

\paragraph{TF1.} 

For this case, using surmised eigenfunctions of the type

\begin{equation} \label{eq:12}
    \Phi_{1} = e^{-\gamma \cos^{2}(x)}\cos^{\eta}(x)f(z)
\end{equation}
with $z=\cos^2(x)$, we find the CHE
\begin{equation} \label{eq:13}
    \frac{d^{2}f}{dz^{2}} + \left[ - 2\gamma + \frac{\eta+\frac{1}{2}}{z} + \frac{\frac{1}{2}}{z-1} \right] \frac{df}{dz} + \left[ \frac{ 
    \frac{1}{4}\left( E - \eta - 2 \gamma \left( 2\eta + 1 \right) \right)}{z} - \frac{
    \frac{1}{4}\left( E - \eta - V_{1} + 2\gamma \left( 2\gamma + 1 \right) \right)}{z-1} \right]f = 0
\end{equation}
and $V_{1} = 8 \gamma [N+ 1 + (\eta + \gamma-1)/2]$, the same expression as in the case of the hyperbolic potential.
For $N = 0$, the energy is $E_0=\eta + 2\gamma( 2\eta + 1 )$;
for $N = 1$, we find
$E_{\pm} = 3\eta + 6\gamma + 4\gamma\eta +2 \pm \big[\left( \eta + 1 \right)^{2} + 4\gamma \left( \gamma - \eta \right)\big]^{1/2}$, with
$z_{1,1}=-2(2\eta+1)/\left(-E_{\pm}+\eta+2\gamma(2\eta+1)\right)$. In the case $N=2$, one obtains (using Mathematica) the eigenenergies $E_{0} = 35.875$, $E_{2} = 54.000$, and $E_{4} = 68.124$, with the three pairs of roots $(z_{2,1},z_{2,2})$ = $(1.124,2.009)$, $(0.388,1.612)$, and $(0.294,0.823)$.

\medskip

\paragraph{TF2.} 
For the odd eigenfunctions, we propose
\begin{equation} \label{eq:14}
    \Phi_{2} = e^{-\gamma \cos^{2}(x)}\cos^{\eta}(x)\sin(x)f(x)
\end{equation}
and find that $V_{1} = 8 \gamma [N+1 + (\eta + \gamma)/2]$, and the CHE
{\small
\begin{equation} \label{eq:15}
    \frac{d^{2}f}{dz^{2}} + \left[ - 2\gamma + \frac{\eta+\frac{1}{2}}{z} + \frac{\frac{3}{2}}{z-1} \right] \frac{df}{dz} + \left[ \frac{ 
    \frac{1}{4}\left( E - 3\eta - 2 \gamma \left( 2\eta + 1 \right)- 1 \right)}{z} - \frac{
    \frac{1}{4}\left( E- 3\eta - V_{1} + 2\gamma \left( 2\gamma + 3 \right)- 1  \right)}{z-1} \right]f = 0~.
\end{equation}
}
For $N = 0$, the energy is $E_0 = \eta + (2\gamma +1)(2\eta + 1)$;
for $N = 1$,
$E_{\pm} = 5 + 5\eta + 6\gamma + 4\gamma\eta \pm 2\big[\left(\eta + 2 \right)^{2} + 4\gamma \left( \gamma - \eta + 1 \right)\big]^{1/2}$, with
$z_{1,1} = - 2 \left( 2\eta + 1 \right)/\left(-E_{\pm} + 2\gamma \left( 2\eta + 1 \right) + 3\eta + 1\right)$. In the case $N=2$,
after solving numerically the system of equations for $\eta=\gamma=2$, the three eigenvalues are $E_{1} = 38.449$, $E_{3} = 61.916$ and $E_{5} = 84.635$ and the three pairs of roots are $(z_{2,1},z_{2,2})$ = $(1.368,2.417)$, $(0.335,1.983)$, and $(0.663,0.235)$.
The analytical and numerically found eigenvalues are displayed in Table \ref{table:3}, for the cases with $\gamma = 2$ and $\eta = 2$, and for
$N = 0$, $1$, and $2$.  Compared to the results in
Table \ref{table:1}, one can notice in Table \ref{table:3} the eigenvalues of opposite sign for the analytically found portion of the spectrum as
a consequence of anti-isospectrality.

\begin{table}[htp!]
\begin{center}
\begin{tabular}{|c|c|c|c|c|c|c|}
\hline
& \multicolumn{3}{c|}{{\em TF1}} & \multicolumn{3}{c|}{{\em TF2}} \\
\hline
  & $N$ = 0 & $N$ = 1 & $N$ = 2 & $N$ = 0 & $N$ = 1 & $N$ = 2 \\
\hline
$E_{0}$ & $\boldsymbol{22.000}$ & $\boldsymbol{30.000}$ & $\boldsymbol{35.875}$
 & $26.400$ & $33.098$ & $38.429$ \\
$E_{1}$ & $23.394$ & $30.247$ & $35.921$ & $\boldsymbol{27.000}$ & $\boldsymbol{33.202}$ & $\boldsymbol{38.449}$ \\
$E_{2}$ & $30.368$ & $\boldsymbol{42.000}$ & $\boldsymbol{54.000}$
 &  $35.979$ & $48.088$ & $59.580$ \\
$E_{3}$ &  $38.656$ & $48.088$ & $57.421$
 &  $43.351$ & $\boldsymbol{52.798}$ & $\boldsymbol{61.915}$ \\
$E_{4}$ & $49.195$ & $58.331$ & $\boldsymbol{68.124}$
 &  $53.703$ & $63.119$ & $73.404$ \\
$E_{5}$ & $61.911$ & $70.764$ & $79.935$
 & $66.299$ & $75.310$ & $\boldsymbol{84.635}$ \\
$E_{6}$ & $76.716$ & $85.383$ & $94.290$
 & $81.020$ & $89.806$ & $98.841$ \\
$E_{7}$ & $93.576$ & $102.113$ & $110.837$
 & $97.822$ & $106.451$ & $115.273$ \\
$\vdots$ & $\vdots$ & $\vdots$ & $\vdots$
 & $\vdots$ & $\vdots$ & $\vdots$ \\
\hline
\end{tabular}
\caption{Exact (bold type) and numerical eigenvalues for the trigonometric potential (\ref{eq:2}), for even {\em TF1} and odd {\em TF2}
eigenfunctions.} 
\label{table:3}
\end{center}
\end{table}

\section{Matching terms in the CHE}\label{eheun} 

We shall compare now the analytical solutions found in Section~\ref{poli}, with those found matching all coefficients
in equations (\ref{eq:4}), (\ref{eq:7}), (\ref{eq:9}), and (\ref{eq:11}), for the hyperbolic cases, with the coefficients of the standard CHE.
The general form of the CHE reads \citep[]{downing:2013,fiziev:2010}

\begin{equation} \label{eq:16}
    \frac{d^{2}H(z)}{dz^{2}} + \left( \alpha + \frac{1 + \beta}{z} + \frac{1 + \gamma^{\ast}}{z-1} \right) \frac{dH(z)}{dz} +
    \left( \frac{\mu}{z} + \frac{\nu}{z-1} \right) H(z) = 0
\end{equation}
and possesses the solution
\begin{equation} \label{eq:17}
    H_{C} \left( \alpha,\beta,\gamma^{\ast},\delta,\eta^{\ast};z \right) = \sum^{\infty}_{i=0} v_{N}
    \left( \alpha,\beta,\gamma^{\ast},\delta,\eta^{\ast} \right) z^{i}~,
\end{equation}
where
\begin{eqnarray}
    \delta &=& \mu + \nu - \frac{\alpha}{2} \left( \beta + \gamma^{\ast} + 2 \right) \label{eq:18}\\
    \eta^{\ast} &=& -\mu +\frac{\alpha}{2} \left( \beta + 1 \right) - \frac{1}{2} \left( \beta + \gamma^{\ast} + \beta \gamma^{\ast} \right) \label{eq:19}
\end{eqnarray}
and the coefficients $v_{N}$ are given by the three-term recurrence relation
\begin{equation} \label{eq:20}
    A_{N}v_{N} = B_{N}v_{N-1} + C_{N}v_{N-2}~, 
\end{equation}
with `initial conditions' $v_{-1}=0,\,v_{0} = 1$ and
\begin{eqnarray}
    A_{N} &=& 1 + \frac{\beta}{N} \label{eq:21} \\
    B_{N} &=& 1 + \frac{1}{N} \left( \beta + \gamma^{\ast} - \alpha - 1 \right) - \frac{1}{2N^{2}}
   \bigg[\left( \beta + \gamma^{\ast}- \alpha \right) -2\eta^{\ast}-\beta(\gamma^{\ast}-\alpha) \bigg] \label{eq:22} \\
    C_{N} &=& \frac{\alpha}{N^{2}} \left(\frac{\beta + \gamma^{\ast}}{2}+\frac{\delta}{\alpha} + N - 1 \right)~. \label{eq:23}
\end{eqnarray}

To reduce a confluent Heun function to a confluent Heun polynomial of degree $N$, we need two termination conditions that must be satisfied simultaneously:

\begin{equation} \label{eq:24}
	\mu + \nu + N\alpha = 0
\end{equation}
and the tridiagonal determinant condition, $\Delta_{N+1} \left( \mu \right) = 0$, which is
\begin{equation} \label{eq:25}
    \left| \begin{matrix}
    \mu - q_{1} & \left( 1 + \beta \right) & 0 & \ldots & 0 & 0 & 0 \\
    N\alpha & \mu - q_{2} + \alpha & 2 \left( 2 + \beta \right) & \ldots & 0 & 0 & 0 \\
    0 & \left( N - 1 \right) \alpha & \mu - q_{3} + 2 \alpha & \ldots & 0 & 0 & 0 \\
    \vdots & \vdots & \vdots & \ddots & \vdots & \vdots & \vdots \\
    0 & 0 & 0 & \ldots & \mu - q_{N-1} + \left( N - 2 \right)\alpha & \left( N - 1 \right)\left( N - 1 + \beta \right) & 0 \\
    0 & 0 & 0 & \ldots & 2\alpha & \mu - q_{N} + \left( N - 1\right)\alpha & N \left( N + \beta \right) \\
    0 & 0 & 0 & \ldots & 0 & \alpha & \mu - q_{N+1} + N\alpha \\
    \end{matrix}
    \right| = 0~,
\end{equation}
where $q_{n} = \left( n - 1 \right)\left( n + \beta + \gamma^{\ast} \right)$.  For comparison, we also need to set $z=\cosh^{2}(x)$.

\subsection{The hyperbolic case}

\paragraph{TF1.}  

We compare the coefficients in eq.~(\ref{eq:4}) to those of the Heun eq.~(\ref{eq:16}) to obtain
\begin{align*}
    \alpha = - 2 \gamma, \ \beta = \eta - \frac{1}{2}, \ \gamma^{\ast} = -\frac{1}{2}, \ \mu = -\frac{E + \eta + 2\gamma \left( 2\eta + 1 \right)}{4}, \ 
    \nu = \frac{E + \eta - V_{1} - 2\gamma \left( 2\gamma + 1 \right)}{4}~.
\end{align*}
With the usage of equations (\ref{eq:18}), (\ref{eq:19}), and (\ref{eq:24}), one finds $\delta = \gamma \left( 2N + \eta + 1 \right)$, $\eta^{\ast} = \frac{1}{8} \left( 2E + 3 \right)$, and $V_{1} = -4 \gamma \left(2N + 1 + \gamma +\eta\right)$.
The wave function reads
\begin{equation} \label{eq:26}
    \psi(x) = e^{-\gamma \cosh^{2}(x)} \cosh^{\eta}(x) H_{C} \left( \alpha,\beta,\gamma^{\ast},\delta,\eta^{\ast}; \cosh^{2}(x)\right)~.
\end{equation}

In the case $N = 0$, one obtains $ \Delta_{1} = \mu - q_{1} = 0$, $q_{1} = 0$, and $\mu = 0$.
Therefore $E_0 = - \eta - 2\gamma( 2\eta + 1)$, as found above. For $N=1$, we find that
\begin{equation*}
    \Delta_{2} = \left| \begin{matrix}
    \mu - q_{1} & 1 + \beta \\
    \alpha & \mu - q_{2} + \alpha
    \end{matrix}
    \right| = 0, \quad \mathrm{with} \ q_{1} = 0 \ \mathrm{and} \ q_{2} = \beta + \gamma^{\ast}+2=\eta+1~,
\end{equation*}
then,
\begin{equation*}
    \mu^{2} - \left( \eta +1 + 2\gamma \right) \mu + 2\gamma \left( \eta + 1/2 \right) = 0~,
    \ \mathrm{where} \ \mu = -\frac{1}{4} \left( E + \eta + 2\gamma \left( 2\eta + 1 \right) \right)~,
\end{equation*}
rendering $E_{\pm} = - \left( 3\eta + 2 \right) \left( \gamma + 1 \right) - \gamma \left( \eta + 4 \right) \pm 2 \big[\left( \eta + 1 \right)^{2} + 4\gamma \left( \gamma - \eta \right)\big]^{1/2}$, and the expansion coefficient
$v_{1} = (E_{\pm} + 2\gamma + \eta + 4\gamma\eta)/2 \left( 2\eta + 1 \right)=\gamma+(E_{\pm}+ \eta)/2 \left( 2\eta + 1 \right)$, which is just $-1/z_{1,1}$ in the polynomial expansion in Section \ref{ssec21}.

Meanwhile, for $N = 2$ we obtain

\begin{equation*}
\Delta_{3} = \left| \begin{matrix}
\mu - q_{1} & 1 + \beta & 0 \\
2\alpha & \mu - q_{2} + \alpha & 2\left( 2+\beta \right) \\
0 & \alpha & \mu - q_{3} + 2\alpha
\end{matrix}
\right| = 0, \ \mathrm{where} \ q_{1} = 0, \ q_{2} =\beta + \gamma^{\ast}+2, \ \mathrm{and} \ q_{3} = 2 \left(\beta + \gamma^{\ast}+3 \right)~,
\end{equation*}
or $q_{1} = 0, \, q_{2} =\eta+1, \, \mathrm{and} \, q_{3} =2\eta+4$ in terms of the parameter $\eta$ of the potential;
so the determinant equation is
\begin{equation*}
    \mu^3 - \left( 6\gamma + 3\eta + 5 \right)\mu^2 + 2\left( \eta^{2} + 8\eta\gamma + 4\gamma^{2} + 3\eta + 10\gamma + 2 \right)\mu - 4\gamma \left( 2\eta + 1 \right) \left( 2\gamma + \eta + 2 \right) = 0~,
\end{equation*}
where $\mu = -\left( E + \eta + 2\gamma \left( 2\eta + 1 \right) \right)/4$.
The expansion coefficients $v_1$ and $v_2$ turn out to be
\[
    v_1 = \gamma + \frac{E + \eta}{2 \left( 2\eta + 1 \right)} \ , \ \
    v_2= \frac{4\gamma}{2\eta+3}+\frac{E + 4\gamma \eta +10\gamma + 5\eta + 4 
    }{4 
    \left( 2\eta + 3 \right)}v_1~.
\]
The numerical eigenvalues correspond to those found in Section~\ref{ssec21}.

\paragraph{TF2.} 

For the odd functions (\ref{eq:7}), matching the terms with those of the standard Heun equation~(\ref{eq:16}), we obtain
\begin{align*}
    \alpha = - 2\gamma, \ \beta = \eta - \frac{1}{2}, \ \gamma^{\ast} = \frac{1}{2}, \ \mu = -\frac{E + 3\eta + 1 + 2\gamma \left( 2\eta + 1 \right)}{4}, \ 
    \nu = \frac{E + 3\eta + 1 - V_{1} - 2\gamma \left( 2\gamma + 3 \right)}{4}~.
\end{align*}
Hence, we find that $\delta = \gamma \left( \eta + 2 + 2N \right)$, $\eta^{\ast} = \frac{1}{8} \left( 2E + 3 \right)$ and $V_{1} = -4 \gamma \left( \gamma + \eta + 2N + 2 \right)$. As for the wave eigenvalues and eigenfunctions, when $N = 0$, we find $E_0 = -\eta - (2\gamma+1)( 2\eta + 1)$.

For the case $N=1$, the determinant is
\begin{equation*}
    \Delta_{2} = \left| \begin{matrix}
    \mu - q_{1} & 1 + \beta \\
    \alpha & \mu - q_{2} + \alpha
    \end{matrix}
    \right| = 0, \quad \mathrm{where} \ q_{1} = 0 \ \mathrm{and} \ q_{2} = 2 + \beta + \gamma^{\ast}\tcm{=\eta+2}~,
\end{equation*}
then,
\begin{equation*}
    \mu^{2} - \left(2\gamma +\eta + 2 \right)\mu + \gamma \left( 2\eta + 1\right) = 0~,
    \ \mathrm{where} \ \mu = -\frac{1}{4} \left( E + 1 + 3\eta + 2\gamma\left( 2\eta + 1 \right) \right)
\end{equation*}
rendering $E_{\pm} = -[5 + 5\eta + 6\gamma + 4\gamma\eta] \pm 2 \big[\left( \eta + 2 \right)^{2} + 4\gamma \left( \gamma - \eta + 1 \right)\big]^{1/2}$. The expansion coefficients $v_{1} = (E_{\pm} + 2\gamma + 3\eta + 1 + 4\gamma\eta)/2 \left( 2\eta + 1 \right)$ are the negative inverse of the roots found in
Section \ref{ssec21}, and the eigenvalues coincide with those found therein.

Moving to the $N = 2$ case, the zero determinant condition is

\begin{equation*}
\Delta_{3} = \left| \begin{matrix}
\mu - q_{1} & 1 + \beta & 0 \\
2\alpha & \mu - q_{2} + \alpha & 2\left( 2+\beta \right) \\
0 & \alpha & \mu - q_{3} + 2\alpha
\end{matrix}
\right| = 0, \ \mathrm{where} \ q_{1} = 0, \ q_{2} = 2 + \beta + \gamma^{\ast} \ \mathrm{and} \ q_{3} = 2 \left( 3 + \beta + \gamma^{\ast} \right)~,
\end{equation*}
(or $q_1=0, q_2=\eta+2, q_3=2(\eta+3)$) which gives
\begin{equation*}
    \mu^{3} - \left( 6\gamma +3\eta + 8 \right)\mu^{2} + 2\left( \eta^{2} + 8\eta\gamma + 4\gamma^{2} + 5\eta + 14\gamma + 6 \right)\mu - 4\gamma \left( 2\eta + 1 \right) \left(2\gamma +\eta +3\right) =0~,
\end{equation*}
where $\mu = -\frac{1}{4} \left( E + 1 + 3\eta + 2\gamma\left( 2\eta + 1 \right) \right)$.
In this case, the eigenvalues are found numerically, while the expansion coefficients have the form
$$
v_{1} = \gamma + \frac{1}{2} +\frac{E + \eta}{2 \left( 2\eta + 1 \right)}~,\, 
v_{2} = \frac{4\gamma}{2\eta+3}+\frac{E + 4\gamma\eta +10\gamma +7\eta + 9
}{4 \left( 2\eta + 3 \right)}v_1~.
$$

\subsection{The trigonometric case}
Let us now look at the solutions for the trigonometric potential (\ref{eq:2}).

\paragraph{ TF1.}  

Comparing eq.~(\ref{eq:13}) to the CHE (\ref{eq:16}), we obtain the followings values
\begin{align*}
    \alpha = - 2 \gamma, \ \beta = \eta - \frac{1}{2}, \ \gamma^{\ast} = -\frac{1}{2}, \ \mu = \frac{E - \eta - 2\gamma \left( 2\eta + 1 \right)}{4}, \ 
    \nu = -\frac{E - \eta - V_{1} + 2\gamma \left( 2\gamma + 1 \right)}{4}
\end{align*}
and using $\mu$ and $\nu$ from eqs.~(\ref{eq:18}), (\ref{eq:19}), and (\ref{eq:24}), we obtain
$\delta = \gamma \left( \eta + 1 + 2N \right)$, $\eta^{\ast} = \frac{1}{8} \left( -2E + 3 \right)$ and $V_{1} = 4\gamma \left(2N + 1 + \gamma + \eta\right)$, as before.
This leads for $N = 0$ to $\Delta_{1} = \mu - q_{1} = 0$ and $q_{1} = 0$; therefore $\mu = 0$, and so  $E_0 = \eta + 2\gamma \left( 2\eta + 1 \right)$.
For $N = 1$, with $q_{1} = 0$ and $q_{2} = 2 + \beta + \gamma^{\ast}$ $(=\eta+1)$, the corresponding determinant condition is
$\mu^{2} - \left( 2 + \beta + \gamma^{\ast} - \alpha \right) \mu - \alpha \left( 1 + \beta \right) = 0$,
where $\mu = \frac{1}{4} \left(E - \eta - 2\gamma \left( 2\eta + 1 \right) \right)$, rendering
$E_{\pm}= 3\eta + 6\gamma + 2 + 4\gamma\eta \pm 2 \big[\left( \eta + 1 \right)^{2} + 4\gamma \left( \gamma - \eta \right)\big]^{1/2}$, and the expansion coefficient $v_{1} = \gamma+(-E+\eta)/(2 \left( 2\eta + 1\right)$.
In the case of $N = 2$, the expansion coefficients are
$$
v_{1} =\gamma+ \frac{-E + \eta}{2 \left( 2\eta + 1 \right)}~, \quad 
v_{2} = \frac{4\gamma}{2\eta+3}+\frac{-E + 4\gamma\eta +10\gamma +5\eta + 4}{4 \left( 2\eta + 3 \right)}v_1~.
$$

\medskip

\paragraph{TF2.} 

For odd solutions, comparing eq.~(\ref{eq:15}) to eq.~(\ref{eq:16}), we find that
\begin{equation*}
    \alpha = - 2 \gamma, \ \beta = \eta - \frac{1}{2}, \ \gamma^{\ast} = \frac{1}{2}, \ \mu = \frac{E- 3\eta- 2\gamma \left( 2\eta + 1 \right)-1}{4}, \ 
    \nu = -\frac{E- 3\eta- V_{1} + 2\gamma \left( 2\gamma + 3 \right)-1}{4}~.
\end{equation*}
Together with $\delta = \gamma \left( \eta + 2 + 2N \right)$, $\eta^{\ast} = \frac{1}{8} \left( -2E + 3 \right)$, we get $V_{1} = 4\gamma \left(2N + 2 + \gamma + \eta\right)$.
Thus, for $N = 0$, we find $E_0 = \eta + (2\gamma+1)(2\eta +1)$; for $N = 1$, we find
$E_{\pm} = 5 + 5\eta + 6\gamma + 4\gamma\eta \pm 2 \big[\left( \eta + 2 \right)^{2} + 4\gamma \left( \gamma - \eta + 1 \right)\big]^{1/2}$,
with $v_{1} = (-E_{\pm} + 2\gamma + 3\eta + 4\gamma\eta + 1)/2 \left( 2\eta + 1 \right)$,
and for $N = 2$, the expansion coefficients are
$$
v_{1} =\gamma + \frac{1}{2}+ \frac{-E +\eta}{2(2\eta+1)}~,\quad
v_{2} = \frac{4\gamma}{2\eta+3}+\frac{-E + 4\gamma\eta +10\gamma +7\eta + 9}{4 \left( 2\eta + 3 \right)}v_1~.
$$
and solve for the energies numerically to find the same values as in Section \ref{ssec22}.

\section{The Lie algebra procedure} \label{liealg}

We now come to seeking the solutions through the Lie algebra approach. For this goal, we shall follow the work of Finkel {\em et al} \citep[]{finkel:1999}, where they solve the Lie algebra for the simpler case of the Razavy potential.

\subsection{The hyperbolic case}\label{ssec41}

We begin by writing the potential function (\ref{eq:1}) in the form
\begin{equation} \label{eq:27}
    V(x;\gamma,\eta,M) = 4\gamma^{2}\cosh^{4}(x) - 4\gamma \left( \eta + \gamma + M \right) \cosh^{2}(x) + \eta \left( \eta-1 \right)\tanh^{2}(x)~,
\end{equation}
where $V_1(\gamma,\eta,M)=- 4\gamma \left( \eta + \gamma + M \right)$, and the parameter $M$ for all four trial functions in Section
\ref{ssec21} is given in Table \ref{table:4}.
We use the potential function for the QES problem studied by Finkel,
\begin{equation} \label{eq:28}
	{\cal V}(x) = A\cosh^{2}(\sqrt{\kappa}x) + B\cosh(\sqrt{\kappa}x) + C\coth(\sqrt{\kappa}x)\csch(\sqrt{\kappa}x) + D\csch^{2}(\sqrt{\kappa}x)~.
\end{equation}
Using $\kappa = 4$, we match the equations (\ref{eq:27}) and (\ref{eq:28}) obtaining $A = \gamma^{2}$, $B = -2 \gamma \left( \eta + M \right)$ and $D = -C = 2\eta \left( 2\eta - 1 \right)$ and some constant which is unnecessary during our procedure. By means of these constants we obtain the gauge transformation for the hyperbolic potential \citep[]{gonzalez-lopez:1993}.

\begin{table}[htp!]
\begin{center}
\begin{tabular}{|c|c|c|c|}
\hline
 & $M$ & $\sigma$ & $\mathrm{Eigenfunctions}$ \\
\hline
{\em TF1} & 2$N$ + 1 & 2$\eta$ & $\Psi_{1} = e^{-\gamma\cosh^{2}(x)}\cosh^{\eta}(x)\,f(\cosh^2(x))$ \\
\hline
{\em TF2} & 2$N$ + 2 & 2$\eta$ + 1 & $\Psi_{2} = e^{-\gamma\cosh^{2}(x)}\cosh^{\eta}(x)\sinh(x)\,f(\cosh^2(x))$ \\
\hline
{\em TF3} & 2$N$+2 - 2$\eta$ & 1 & $\Psi_{3} = e^{-\gamma\cosh^{2}(x)}\sech^{\eta-1}(x)\,f(\cosh^2(x))$ \\
\hline
{\em TF4} & 2$N$+3 - 2$\eta$ & 2 & $\Psi_{4} = e^{-\gamma\cosh^{2}(x)}\sech^{\eta-1}(x)\sinh(x)\,f(\cosh^2(x))$ \\
\hline
\end{tabular}
\caption{The four types of proposed forms of eigenfunctions for the hyperbolic potential (\ref{eq:1}).
The parameter $\sigma$ appears in the gauge transformation function $\hat{\mu}(z)$.}
\label{table:4}
\end{center}
\end{table}
Setting  $\psi(z;\sigma) = \hat{\mu}(z;\sigma) \hat{\chi}(z;\sigma)$, and using the gauge transformation function in terms of the variable $z=\cosh(2x)$
\begin{equation} \label{eq:29}
    \hat{\mu}(z;\sigma) = \left( z - 1 \right)^{\frac{1}{4} \left( \sigma -\eta + \frac{\eta \left( \eta - 1 \right)}{ \sigma -\eta - 1} \right)} \left( z + 1 \right)^{ \frac{1}{4} \left( \sigma -\eta - \frac{\eta \left( \eta - 1 \right)}{ \sigma -\eta -1} \right)} e^{-\frac{\gamma}{2}z}
\end{equation}
and the $\mathfrak{sl}(2,\mathbb{R})$ operators
\begin{equation*}
    J_{-} = \partial_{z}, \ J_{0} = z \partial_{z} - \frac{N}{2}, \ J_{+} = z^{2}\partial_{z} - Nz~,
\end{equation*}
we find that the gauge Hamiltonian may be written as
\begin{equation} \label{eq:30}
    \hat{H}_{g}(z;\sigma) =
-4J_{0}^{2} + 4J_{-}^{2} + 4\gamma J_{+} - 4 \left(\sigma - \eta + N \right) J_{0} + 4\left(\frac{\eta \left( \eta - 1 \right)}{ \sigma-\eta -1} -\gamma \right)J_{-} + c_{\ast\, h}(\sigma)~,
\end{equation}
where $c_{\ast\, h}(\sigma) = -(N+\sigma -\eta)^2- 2\gamma \left(2N+\sigma -\eta + 1 \right)
+ \eta \left( \eta - 1 \right) - \frac{2\gamma\eta\left( \eta - 1 \right)}{ \sigma-\eta - 1}$.

Following the Lie algebraic method, we only need to look for the orthogonal polynomial solutions
\begin{equation} \label{eq:31}
    \hat{\chi}(z;\sigma) = \sum_{k = 0}^{\infty} \frac{\left( \frac{\sigma}{2} + \frac{\eta \left( 2\eta - \sigma \right)}{2 \left( \sigma - \eta -1\right)} + k \right)\,!}{2^{k}\left( \sigma + \frac{\eta \left( 2\eta - \sigma \right)}{ \sigma -\eta -1} + 2k \right)\,!} \frac{\left( z + 1 \right)^{k}}{k\,!}\hat{P}_{k}~,
\end{equation}
where the three-term recurrence relation
\begin{equation}\label{eq:32}
    \hat{P}_{k+1} = (E - b_{k}(\sigma)) \hat{P}_{k} - a_{k}(\sigma) \hat{P}_{k-1}, \quad k \geq 0
\end{equation}
is satisfied for
\begin{align*}
    a_{k}(\sigma) &= 16\gamma k \left( k - N - 1 \right)\left( 2k - 1 + \sigma - \eta + \frac{\eta \left( \eta - 1 \right)}{\sigma - \eta -1} \right) \\
    b_{k}(\sigma) &= -4k \left( \sigma - \eta + k + 2\gamma \right) - 2\gamma \left(\sigma -\eta + 1 \right) + \eta \left( \eta - 1 \right) - \frac{2\gamma\eta\left( \eta - 1 \right)}{\sigma- \eta -1} - \left(\sigma - \eta \right)^{2}~.
\end{align*}

Eigenvalues are found by setting $\hat{P}_{k+1} = 0$.  Two kinds of solutions are found in this way, which can be divided in two more categories to include negative $\eta$'s in Table~\ref{table:4}. The corresponding eigenfunctions are given by
\begin{equation} \label{eq:33}
    \psi(z;\sigma) = \left( z + 1 \right)^{\frac{1}{4} \left(\sigma -\eta  + \frac{\eta \left( \eta - 1 \right)}{\sigma-\eta - 1} \right)} \left( z - 1 \right)^{ \frac{1}{4} \left(\sigma -\eta - \frac{\eta \left( \eta - 1 \right)}{\sigma-\eta -1} \right)} e^{-\frac{\gamma}{2}z} \sum_{k = 0}^{\infty} \frac{\left( \frac{\sigma}{2} + \frac{\eta \left( 2\eta - \sigma \right)}{2 \left(\sigma- \eta -1 \right)} + k \right)\,!}{2^{k}\left( \sigma + \frac{\eta \left( 2\eta - \sigma \right)}{\sigma - \eta -1} + 2k \right)\,!} \frac{\left( z + 1 \right)^{k}}{k\,!}\hat{P}_{k}~.
\end{equation}

We now apply these general results to the first two cases of Table~\ref{table:4}, for the hyperbolic case.

\paragraph{ TF1.} 

The first case is when $\sigma = 2\eta$ and $M = 2N +1$,
and the coefficients in the three-term recurrence relation (\ref{eq:32}) become
\begin{align*}
    a_{k} &= 16\gamma k \left( k - N - 1 \right)\left( 2k - 1 + 2\eta \right) \\
    b_{k} &= -4k \left( \eta + k + 2\gamma \right) - 2\gamma \left( 2\eta + 1 \right) - \eta~.
\end{align*}

The even eigenfunctions are
\begin{equation*}
    \psi(z)\equiv \psi(z;2\eta) = \left( z+1 \right)^{\frac{\eta}{2}} e^{- \frac{\gamma}{2} z} \sum_{k = 0}^{\infty} \frac{\left( \eta + k \right)\,!}{2^{k}\left( 2\eta + 2k \right)\,!} \frac{\left( z + 1 \right)^{k}}{k\,!}  \hat{P}_{k}~.
\end{equation*}

For example, assuming monic polynomials, $\hat{P}_{0}(E) = 1$,
 we obtain the following results. For $N = 0$, from $\hat{P}_{1}(E)\equiv \left( E - b_{0} \right) \hat{P}_{0} - a_{0} \hat{P}_{-1} = 0$, the eigenvalue is $E_0 = - \eta - 2\gamma( 2\eta + 1)$.
For $N=1$, we have
\begin{align*}
    \hat{P}_{1} &= \left( E + 2\gamma \left( 2\eta + 1 \right) + \eta \right) \hat{P}_{0}  \\
    \hat{P}_{2} &= \left( E + 2\gamma \left( 2\eta + 5 \right) + 5\eta + 4 \right) \hat{P}_{1} + 16\gamma \left( 2\eta + 1 \right) \hat{P}_{0}~
\end{align*}
and with the condition $\hat{P}_{2} = 0$, we find the eigenvalues $E_{\pm} = -3\eta - 6 \gamma - 2 - 4\gamma\eta \pm 2[\left( \eta + 1 \right)^{2} + 4\gamma \left( \gamma - \eta \right)]^{1/2}$.
Furthermore, in the case with $N = 2$, we have
\begin{align*}
    \hat{P}_{1} &= E + 2\gamma \left( 2\eta + 1 \right) + \eta \\
    \hat{P}_{2} &= \left( E + 2\gamma \left( 2\eta + 5 \right) + 5\eta + 4 \right) \hat{P}_{1} + 32\gamma \left( 2\eta + 1 \right) \hat{P}_{0} \\
    \hat{P}_{3} &=  \left( E + 2\gamma \left( 2\eta + 9 \right) + 9\eta + 16 \right) \hat{P}_{2} + 32\gamma \left( 2\eta + 3 \right) \hat{P}_{1}\\
\end{align*}
and setting $\hat{P}_{3} =0$, we find the eigenvalues that coincide with those given in Section \ref{ssec21}.

\paragraph{ TF2.} 

For the odd solutions of the second case, with $\sigma = 2\eta + 1$ and $M = 2N + 2$, we have
\begin{align*}
    a_{k} &= 16\gamma k \left( k - N - 1 \right)\left( 2k + 2\eta - 1 \right) \\
    b_{k} &= -4k \left( \eta + 1 + k + 2\gamma \right) - 2\gamma \left( 2\eta + 1 \right) - 3\eta - 1
\end{align*}
and the eigenfunctions are
\begin{equation*}
	\psi(z)\equiv \psi(z;2\eta + 1) = \left( z + 1 \right)^{\frac{\eta}{2}} \left( z - 1 \right)^{\frac{1}{2}}
e^{-\frac{\gamma}{2}z} \sum_{k = 0}^{\infty} \frac{\left( \eta + k \right)\,!}{2^{k}\left( 2\eta +
2k \right)\,!} \frac{\left( z + 1 \right)^{k}}{k\,!}\hat{P}_{k}~.
\end{equation*}

For $N = 0$, we have $\hat{P}_{0} = 1$ and $\hat{P}_{1} = E + 2\gamma \left( 2\eta + 1 \right) + 3\eta + 1$;
therefore, the eigenvalue is $E_0 =-\eta -(2\gamma+1)(2\eta + 1)$. For $N$ = 1, we find
$\hat{P}_{1}= E + 2\gamma \left( 2\eta + 1 \right) + 3\eta + 1$ and
$\hat{P}_{2}=\left( E + 2\gamma \left( 2\eta + 5 \right) + 7\eta + 9  \right)\left( E + 2\gamma \left( 2\eta + 1 \right) + 3\eta + 1 \right) + 16\gamma\left( 2\eta + 1 \right)$, from which we find the eigenvalues
$E_{\pm} = -5 - 5\eta - 6\gamma - 4\gamma\eta \pm 2 [\left( \eta + 2 \right)^{2} + 4\gamma \left( \gamma - \eta + 1 \right)]^{1/2}$.

For $N = 2$, we obtain three eigenvalues as solutions of the equation
\begin{align*}
    & \left( E + 11\eta+ 2\gamma \left( 2\eta + 9 \right) + 25 \right)\left( E+ 7\eta + 2\gamma \left( 2\eta + 5 \right) + 9 \right)\left( E+ 3\eta + 2\gamma \left( 2\eta + 1 \right) + 1 \right) \\
    &+ 32\gamma[\left( 2\eta + 1 \right)\left( E+ 11\eta + 2\gamma \left( 2\eta + 9 \right) + 25 \right) + \left( 2\eta + 3\right)\left(E  + 3\eta+ 2\gamma \left( 2\eta + 1 \right) + 1 \right)] =0~.
\end{align*}

\subsection{The trigonometric case}

We now consider the case of the trigonometric potential function (\ref{eq:2})
\begin{equation} \label{eq:34}
    U(x;\gamma,\eta,M) = -4\gamma^{2}\cos^{4}(x) + 4\gamma \left( \eta + \gamma + M \right) \cos^{2}(x) + \eta \left( \eta - 1 \right)\tan^{2}(x)~,
\end{equation}
where $V_{1}(\gamma,\eta,M) = 4\gamma \left( \eta + \gamma + M \right)$ and Table \ref{table:5} provides $M$ and the eigenfunctions. We shall use as reference the algebra developed by \citep[]{finkel:1996} for the potential function
\begin{equation} \label{eq:35}
    {\cal U}(x) = A\sin^{2}(\sqrt{\kappa}x) + B\sin(\sqrt{\kappa}x) + C\tan(\sqrt{\kappa}x)\sec(\sqrt{\kappa}x) + D\sec^{2}(\sqrt{\kappa}x)
\end{equation}
under the change of variable $x\rightarrow x - \pi/4$ and $\kappa =4$, we match the equations (\ref{eq:34}) and (\ref{eq:35}). Therefore we obtain the values of the constants $A = - \gamma^{2}$, $B = -2\gamma \left( \eta + M \right)$, and $C = D = 2\eta \left(\eta - 1\right)$. Using these constants, we obtain the gauge transformation function \citep[]{gonzalez-lopez:1993}:
\begin{equation} \label{eq:36}
    \hat{\mu}(z;\sigma) = \left( 1 + z \right)^{\frac{1}{4} \left(\sigma -\eta - \frac{\eta \left( \eta - 1 \right)}{\sigma-\eta - 1}\right)}\left(1 - z \right)^{ \frac{1}{4} \left(\sigma -\eta + \frac{\eta \left( \eta - 1 \right)}{\sigma-\eta - 1} \right)} e^{\frac{\gamma}{2}z}
\end{equation}
and find the gauge Hamiltonian
\begin{equation*}
    \hat{H}_{g}(z;\sigma) =
    4J_{0}^{2} -4J_{-}^{2} + 4\gamma J_{+} + 4 \left(\sigma - \eta + N \right) J_{0} - 4\left( \gamma - \frac{\eta \left( \eta - 1 \right)}{\sigma - \eta -1} \right)J_{-} + c_{\ast\, t}(\sigma)~,
\end{equation*}
where $c_{\ast\, t}(\sigma)=-c_{\ast\, h}(\sigma)$.
Then, for the eigenfunctions $\psi(z;\sigma) = \hat{\mu}(z;\sigma) \hat{\chi}(z;\sigma)$, we look for the orthogonal polynomials part
\begin{equation} \label{eq:37}
    \hat{\chi}_{E}(z\sigma) = \sum_{k = 0}^{\infty} \left( -1 \right)^{k} \frac{\left( \frac{\sigma}{2} + \frac{\eta \left( 2\eta - \sigma \right)}{2 \left( \sigma -\eta -1 \right)} + k \right)\,!}{2^{k} \left( \sigma + \frac{\eta \left( 2\eta - \sigma \right)}{ \sigma - \eta -1} + 2k \right)\,!} \frac{\left(1-z \right)^{k}}{k\,!} \hat{P}_{k}
\end{equation}
with the same three-term recurrence relation given by (\ref{eq:32}), same $a_k(\sigma)$, but opposite sign $b_k(\sigma)$.
%
%

\begin{table}[htp!]
\begin{center}
\begin{tabular}{|c|c|c|c|}
\hline
 & M & $\sigma$ & $\mathrm{Trigonometric \ transformation}$ \\
\hline
1T & 2N + 1 & 2$\eta$ & $\Phi_{1} = e^{-\gamma\cos^{2}(x)}\cos^{\eta}(x)f(x)$ \\
\hline
2T & 2N + 2 & 2$\eta$ + 1 & $\Phi_{2} = e^{-\gamma\cos^{2}(x)}\cos^{\eta}(x)\sin(x)f(x)$ \\
\hline
3T & 2N+2 - 2$\eta$ & 1 & $\Phi_{3} = e^{-\gamma\cos^{2}(x)}\sec^{\eta}(x)\cos(x)f(x)$ \\
\hline
4T & 2N+3 - 2$\eta$ & 2 & $\Phi_{4} = e^{-\gamma\cos^{2}(x)}\sec^{\eta}(x)\cos(x)\sin(x)f(x)$ \\
\hline
\end{tabular}
\caption{The same as in the previous Table for the trigonometric potential.}
\label{table:5}
\end{center}
\end{table}

\paragraph{ TF1.}

For even functions in this case, we have that $\sigma = 2\eta$ and $M = 2N +1$.

We solve again for the three cases $N$ = 0, $1$, and $2$.
For $N$ = $0$, the identity $\hat{P}_{1}=0$ gives $E_0 = \eta+2\gamma\left( 2\eta + 1 \right)$.
For $N = 1$, when $\hat{P}_{2} = 0$ we obtain
$\left( E - 2\gamma \left( 2\eta + 5 \right) - 5\eta - 4 \right) \left( E - 2\gamma\left( 2\eta + 1 \right) - \eta \right) + 16\gamma\left( 2\eta + 1 \right) = 0$.
In the case $N = 2$, setting
\begin{align*}
    \hat{P}_{3} = & \left( E - 2\gamma \left( 2\eta + 9 \right) - 9\eta - 16 \right)\left( E - 2\gamma \left( 2\eta + 5 \right) - 5\eta - 4 \right)\left(E - 2\gamma\left( 2\eta + 1 \right) - \eta \right)  \\
    &+ 32\gamma[\left( 2\eta + 1 \right)\left( E - 2\gamma \left( 2\eta + 9 \right) - 9\eta - 16 \right) + 
    \left( 2\eta + 3 \right)\left( E - 2\gamma\left( 2\eta + 1 \right) - \eta \right)]=0~
\end{align*}
enables us to find the corresponding eigenvalues.

\paragraph{ TF2.} 

For the odd solutions in the trigonometric case, we set $\sigma = 2\eta + 1$ and $M = 2N + 2$.

\noindent For $N = 0$, the condition $\hat{P}_{1} = 0$ gives the eigenvalue
$E_0 = \eta +(2\gamma+1)(2\eta + 1)$.
For $N = 1$,
the condition $\hat{P}_{2} = \left( E - 2\gamma \left( 2\eta + 5 \right) - 7\eta - 9 \right) \left( E - 2\gamma \left( 2\eta + 1 \right) - 3\eta - 1 \right) + 16\gamma\left( 2\eta + 1 \right)=0$ gives the desired energy eigenvalues, and for $N = 2$, one can obtain the energy eigenvalues from
\begin{align*}
    \hat{P}_{3} &= \left( E - 2\gamma \left( 2\eta + 9 \right) - 11\eta - 25 \right)\left( E - 2\gamma \left( 2\eta + 5 \right) - 7\eta - 9 \right)\left(E - 2\gamma \left( 2\eta + 1 \right) - 3\eta - 1 \right) \\
    &+ 32\gamma[\left( 2\eta + 1 \right)\left( E - 2\gamma \left( 2\eta + 9 \right) - 11\eta - 25 \right) +\left( 2\eta + 3 \right)\left( E - 2\gamma \left( 2\eta + 1 \right) - 3\eta - 1 \right)]=0~.
\end{align*}
%

\section{Conclusions}

Using a three-term hyperbolic potential, not previously discussed in the literature, and its anti-isospectral counterpart, we have shown
the equivalence of the three preferred algebraic procedures to find the exact solutions of QES one-dimensional Schr\"odinger problems, solving three examples for the even and odd solutions in each case.
In some cases, the analytical values are more directly found by one of the procedures than the others, as was shown in the Lie algebra approach, in which the Hamiltonian has been written in a well-settled operatorial form using the gauge functions.


\bigskip
\bigskip

%
%
%
%
%
%

\bigskip

\section*{Acknowledgments}
The first author thanks CONACyT-Mexico for a doctoral fellowship.



\end{document}